\documentclass[aps,prd,preprint,tightenlines,superscriptaddress,nofootinbib]{revtex4}
\usepackage{epsfig}
\usepackage{graphicx}
\usepackage{psfrag}
\usepackage{amsmath,amssymb}
\usepackage{colordvi}
\usepackage{amsfonts}
\usepackage{enumerate}
\usepackage{slashed}
\usepackage{color}
\usepackage{xcolor}

\begin{document}

\title{Probing Parton Orbital Angular Momentum \\in Longitudinally Polarized Nucleon}

\author{Xiangdong Ji}
\affiliation{INPAC, Department of Physics, and Shanghai Key Lab for Particle Physics and Cosmology,
Shanghai Jiao Tong University, Shanghai, 200240, P. R. China}
\affiliation{Center for High-Energy Physics, Peking University, Beijing, 100080, P. R. China}
\affiliation{Maryland Center for Fundamental Physics, University of Maryland, College Park, Maryland 20742, USA}
\author{Xiaonu Xiong}
\affiliation{Center for High-Energy Physics, Peking University, Beijing, 100080, P. R. China}
\affiliation{Nuclear Science Division, Lawrence Berkeley
National Laboratory, Berkeley, CA 94720, USA}
\author{Feng Yuan}
\affiliation{Nuclear Science Division, Lawrence Berkeley
National Laboratory, Berkeley, CA 94720, USA}
\date{\today}
\vspace{0.5in}
\begin{abstract}

While the total orbital angular momentum (OAM) of a definite quark flavor in a longitudinally-polarized nucleon
can be obtained through a sum rule involving twist-two generalized parton distribution (GPDs),
its distribution as a function of parton momentum in light-front coordinates is more complicated to
define and measure because it involves intrinsically twist-three effects. In this paper,
we consider two different parton OAM distributions. The first is manifestly
gauge invariant, and its moments are local operators and calculable in lattice QCD.
We show that it can potentially be measured through twist-three GPDs. The second is
the much-debated canonical OAM distribution natural in free-field theory and light-cone
gauge. We show the latter in light-cone gauge can also be related to twist-three GPDs as well as quantum
phase-space Wigner distributions, both being measurable in high-energy experiments.

\end{abstract}

\maketitle

\section{Introduction}

The nucleon spin structure is one of the most active research areas in hadronic physics
in recent years~\cite{Boer:2011fh}. A gauge-invariant and frame-independent
approach was put forward in~\cite{Ji:1996ek}, according to which, the nucleon
polarization (either longitudinal or transverse) can be decomposed into
frame-independent quark and gluon contributions,
\begin{eqnarray}
\frac{1}{2}&=&\sum_q J_q+J_g \ ,
\end{eqnarray}
where $J_q$ and $J_g$ can be extracted from the following sum rule,
\begin{eqnarray}
J_{q,g}&=&\frac{1}{2}\int d x x\left(H_{q,g}(x,0,0)+E_{q,g}(x,0,0)\right) \ , \label{spinsumrule}
\end{eqnarray}
where $H_{q,g}$ and $E_{q,g}$ are the relevant twist-two generalized parton
distributions (GPDs) for the quarks and gluons, respectively. The above result, however,
does not seem to provide a simple partonic interpretation for the individual contributions,
which does exist, for example, for the quark helicity contribution $\Delta q(x)$
to the nucleon helicity~\cite{Filippone:2001ux}.

In our recent publications~\cite{Ji:2012sj,transverse}, we have
investigated the parton physics of the spin sum rule from the consideration of
Pauli-Lubanski spin vector and angular momentum (AM) density. We found that
for the transverse polarization, the leading contribution has a simple partonic
interpretation that $J_{q/g}(x)=(x/2)(H_{q/g}(x,0,0)+E_{q/g}(x,0,0))$ are just the quark and gluon
angular momentum densities, whereas the sub-leading effects can be taken
into account by the Lorentz symmetry~\cite{transverse}. In other words, Eq.(2)
can be interpreted as a partonic sum rule
for the transverse polarization. For the longitudinal polarization, on the other hand,
the nucleon helicity naturally receives contributions from the parton helicity
and orbital angular momentum (OAM). The quark and gluon OAM densities in
light-front coordinates are not entirely leading-twist effects and therefore are
difficult to define and measure. In the work of Hoodbhoy et al~\cite{Hoodbhoy:1998yb},
the partonic AM densities were defined starting from the generalized AM tensors.
The OAM distribution was identified  as the difference of the total AM density
and the helicity distribution. A careful examination of the operator
structure indicates that this OAM density contains
extra quark and gluon mixing contribution. To keep the physics simple,
we suggested to define the gauge-invariant quark OAM distribution $L_q(x)$ without
this extra term~\cite{Ji:2012sj}. Then, one can show that $L_q(x)$ is related to
twist-three GPDs, which might be measured directly from the hard exclusive processes in
lepton-nucleon scattering~\cite{belitskydvcs}.

The much-discussed partonic OAM distributions in the literature
have been centered on the canonical AM expression~\cite{Jaffe:1989jz,Hagler:1998kg,Harindranath:1998ve,Bashinsky:1998if}.
This definition is not guage invariant out right, but can be made
so through trivial gauge-invariant extension (GIE)
of the light-cone gauge and light-front coordinates~\cite{Ji:2012sj,Ji:2012gc}.
It can be shown that these distributions can also be related to
twist-three parton distributions~\cite{Hatta:2011ku,Ji:2012sj}.
Meanwhile, recent studies~\cite{Ji:2012sj,Lorce:2011kd,Hatta:2011ku} have also shown that the
quark OAM distributions are connected to the
quantum phase space Wigner distributions~\cite{Belitsky:2003nz}.
These distributions define the correlations of partons
in transverse momentum and transverse coordinate spaces.
The gauge-invariant OAM distribution discussed in the previous paragraph and
the canonical OAM distribution in light-cone gauge
are just the projections of the Wigner distributions with different
choices of the associated gauge links.

For the gluon contribution $J_g$ to the helicity, there is no gauge-invariant decomposition
of the operator into the local ones corresponding to the gluon spin
and OAM~\cite{Ji:1996ek}. However, there is a decomposition in a fixed gauge into
the canonical contributions. To relate them to partonic physics,
a GIE of these contributions can be applied to
the operators in the light-cone gauge, just like that for the quark case.
The GIE procedure provides a practical way to connect
the gluon spin and OAM contributions to physical observables~\cite{Ji:2012sj,Ji:2012gc}.

The paper is organized as follows. In Sec.II, we will review
the definitions of the partonic OAM distributions, we consider both
naturally gauge-invariant approach as well as the canonical
definition in light-cone gauge. We explore the relation between these approaches.
In Sec. III, we analyze the twist-three GPDs and their role in directly
probing the quark and gluon OAM contributions. We also discuss the Wigner distributions
for the quarks and gluons, and their connections to the OAM. We summarize
the results in Sec.~IV.



\section{Definitions of Parton Orbital Angular Momentum}

In this section, we {\it review} and relate the definitions of the
parton OAM distributions in a nucleon with momentum $P^\mu = (E, 0, 0, P)$
and definite helicity or the definite angular momentum $J^z=1/2 ~(\hbar=1)$.
The fact that there are more than one definitions reflects the difficulty of finding
one satisfying all the required properties, namely 1) gauge symmetry,
2) clear physical interpretation, 3) measurability in high-energy scattering.
This difficulty originates possibly from the fact that in
a longitudinally polarized nucleon the OAM of the partons is intrinsically a
twist-three effect~\cite{Ji:2012sj}. In this sense, the
parton structure of the helicity is more complicated than that of the transverse
polarization~\cite{transverse}.

We start with the parton OAM distribution by Hoodbhoy et al.~\cite{Hoodbhoy:1998yb}.
This definition starts from the generalized AM tensor and derives
the parton OAM density from twist-two parton distributions. It
emphasizes the experimental measurability and is gauge invariant. However, the
physical interpretation in partons is complicated. Then we consider
an improved definition by inserting the gauge-invariant OAM into
a tower of twist-two operators. This definition is gauge-invariant
and has a clearer physical meaning. We will call this one
the gauge-invariant OAM distribution $L_q(x)$. However, as we shall see, its measurement
is more difficult as it involves twist-three GPDs.  For the same reason,
its partonic interpretation is not completely straightforward in the presence of the transverse
gluon gauge potential. The definition that has been studied the
most in the literature has been motivated from the canonical OAM without the
transverse gluon potential. It is not manifestly gauge invariant.
In parton physics, this definition can be gauge-fixed in the light-cone gauge
and made gauge invariant through an extension of the concept of gauge invariance~\cite{Ji:2012gc}.
Recent work shows that such light-cone gauge, canonical OAM distribution might be measurable
through twist-three GPDs and Wigner distributions~\cite{Lorce:2011kd,hatta,Ji:2012sj}, although this
is even more difficult to achieve than $L_q(x)$.

In the third subsection, we comment on the relationship between the manifestly gauge-invariant
and the light-cone-gauge-motivated definitions.

\subsection{Gauge-Invariant OAM Distribution}

Hoodbhoy et al. have defined a version of the quark OAM distribution from
the generalized energy-momentum tensor~\cite{Hoodbhoy:1998yb}. Recall
that the angular momentum operator can be obtained from the rank-3 angular
momentum tensor made of the symmetric energy-momentum tensor
$T^{\mu\alpha}$ ~\cite{Jaffe:1989jz,Ji:1996ek},
\begin{equation}
M^{\mu\alpha\beta}=\xi^\alpha T^{\mu\beta}-\xi^\beta T^{\mu\alpha} \ ,
\end{equation}
where $\xi$ is a space-time coordinates, $T^{\mu\alpha}$ can be separated into quark and gluon contributions,
\begin{eqnarray}
T^{\mu\alpha}=T^{\mu\alpha}_q+T^{\mu\alpha}_g \ .
\end{eqnarray}
The quark and gluon components follows from the QCD lagrangian,
\begin{eqnarray}
T^{\mu\alpha}_q&=&\frac{1}{2}\left[\overline\psi\gamma^{(\mu}i\overrightarrow{D}^{\alpha)}\psi
+\overline\psi\gamma^{(\mu}i\overleftarrow{D}^{\alpha)}\psi\right]\nonumber\\
T^{\mu\alpha}_g&=&\frac{1}{4}F^2g^{\mu\alpha}-F^{\mu\rho}{F^{\alpha}}_{\rho} \ ,\label{en}
\end{eqnarray}
where the covariant derivative follows the convention $\overrightarrow {D}^\mu=\partial^\mu+igA^\mu$, $\overleftarrow {D}^\mu=-\partial^\mu+igA^\mu$,
and $F^{\mu\nu}$ is the field strength tensor for the gauge field.
In the above expression, we have neglected the contributions from the gauge-fixing term
in the lagrangian which usually yield vanishing physical matrix elements. From the above,
the quark OAM operator is found to be,
\begin{eqnarray}
 L_q &=& \int d^3\xi ~\overline{\psi}_q\gamma^+ \left(\xi^1 (iD^2) - {{\xi^2 (iD^1)}}\right) \psi_q\  .
\label{quarkang}
\end{eqnarray}
This procedure can be generalized to define the quark OAM distribution in the same
way that the parton momentum distribution follows from generalizing the energy-momentum
tensor to a tower of twist-two operators.


The generalized AM tensors can be defined as
\begin{equation}
  M^{\alpha\beta\mu_1...\mu_n}_q(\xi)
  = \xi^\alpha O_q^{\beta\mu_1...\mu_n}
   - \xi^\beta O_q^{\alpha\mu_1...\mu_n} - (\rm trace) \ ,
\end{equation}
where $O_q^{\beta\mu_1...\mu_n}(\xi)
= \bar \psi \gamma^{(\beta}i\overleftrightarrow{D}^{\mu_1}\cdots i\overleftrightarrow{D}^{\mu_n)}\psi(\xi)$
represents the tower of twist-two operators generalizing the quark energy-momentum tensor ($n=1$),
with all indices symmetrized and traces subtracted. The above operators have angular-momentum-dependent
nucleon matrix elements $J_{qn}$ in a state with polarization vector $S^\mu$(see Eq. (4)
in Ref. \cite{Hoodbhoy:1998yb}). The quark
angular momentum distribution $J_q(x)$ can be defined as
\begin{equation}
     \int dx x^{n-1} J_q(x) = J_{qn} \ ,
\end{equation}
just like the moments of the quark momentum distribution which are the matrix elements of the generalized
energy-momentum tensors. Using the GPDs, it has been shown that
\begin{equation}
    J_q(x) = \frac{x}{2} [q(x)+ E_q(x)] \label{jqx}
\end{equation}
where $q(x)$ is the quark momentum distribution and $E_q(x)$ is one
of the twist-two GPDs.

While the partonic content of the above procedure is simple
in the case of transverse polarization~\cite{transverse}, it does not generate a simple
quark OAM distribution in a fixed helicity state. Indeed, by examining
the matrix element of the component $M^{12+...+}$, one may define
\begin{eqnarray}
    \tilde L_q(x) &=&  J_q(x) - \frac{1}{2} \Delta q(x)  \nonumber \\
            &=& \frac{1}{2} \left[ x(q(x) + E_q(x)) - \Delta q(x)\right] \ ,
\end{eqnarray}
as the quark OAM distribution. However,
the moments of the OAM distribution is related to the matrix elements of the following
operator,
\begin{eqnarray}
    \tilde L_q^{+...+} &=& \frac{1}{n}\int d^3\xi
    \left[\overline{\psi} \gamma^+ (\xi^1 iD^2-\xi^2 iD^1) iD^+\cdots iD^+\psi
     + \cdots \right.\nonumber \\ && \left.+~\overline{\psi} \gamma^+  iD^+ \cdots iD^+(\xi^1 iD^2-\xi^2 iD^1)\psi\right]\nonumber \\
 && +~ \frac{1}{n(n+1)} \int d^3\xi
  \left[ \overline{\psi}\gamma^+ (\xi^1\gamma^2-\xi^2\gamma^1) (igF^{\rho+}\gamma_\rho)
    iD^+ \cdots iD^+\psi + \cdots \right.\nonumber \\
    && \left.   + ~\overline{\psi}\gamma^+
    iD^+ \cdots iD^+ (\xi^1\gamma^2-\xi^2\gamma^1) (igF^{\rho+}\gamma_\rho)\psi \right] \ .
\end{eqnarray}
Apart from the first term that has a physical meaning as generalized OAM operator,
it also contains a term proportional to the gluon field strength $F^{\mu\nu}$, whose
physical origin seems obscure. Apparently, the extra contribution comes from the requirement that the
AM density is a twist-two operator.

Thus, we define proper gauge-invariant OAM distribution in the nucleon helicity state from the
following tower of operators,
\begin{eqnarray}
    L_q^{\mu_1...\mu_{n}} = \frac{1}{n}\sum_{i=0}^{n-1}\int d^3\xi~
    \overline{\psi}(\xi) \gamma^+ iD^{\mu_1}\cdots iD^{\mu_{i}}(\xi^1 iD^2-\xi^2 iD^1) iD^{\mu_{i+1}}\cdots iD^{\mu_n}\psi(\xi) \ .\label{moment}
\end{eqnarray}
Considering the matrix elements
\begin{equation}
  \langle PS|L_q^{+...+}|PS\rangle = \frac{2L_{qn}}{n+1} 2S^+P^+\cdots P^+ (2\pi)^3 \delta^3(0) \ ,
\end{equation}
we define the associated OAM distribution $L_q(x)$,
\begin{equation}
              \int dx L_q(x) x^{n-1}  = L_{qn} \ .
\end{equation}
$L_q(x)$ is gauge invariant and has a simple physical meaning, as it involves only
the insertion of the angular momentum operator $\xi^1 iD^2-\xi^2 iD^1$ into the twist-two generalized
energy-momentum tensors.

However, $L_q^{\mu_1...\mu_n}$ contains both twist-two and -three operators, and
depends on the correlation of transverse distributions in both
the coordinate and momentum spaces as well as the transverse gluon potential. Therefore,
its partonic content is not simple. The fact that OAM is a higher-twist operator
in a longitudinally-polarized nucleon can already be appreciated from the
matrix element of $M^{+12}$, which is a subleading operator in light-front
coordinates~\cite{Ji:2012sj}.

We shall see that $L_q(x)$ naturally follows from the (angular momentum) moment of a Wigner
distribution with the straightline gauge link~\cite{Ji:2012sj}. We will also show in the next section
that $L_q(x)$ is related explicitly to twist-two and -three GPDs, and hence is accessible
experimentally in principle.

\subsection{Canonical OAM Distribution in Light-Cone Gauge}

There are two definitions of quark angular momentum in the literature. Apart
from the gauge-invariant definition~\cite{Ji:1996ek}, there is the version
motivated in free-field theory, expressed in the light-cone coordinates~\cite{Jaffe:1989jz},
\begin{eqnarray}
J^3 = \int d^3{{\vec{\xi}}}\left[ \frac{1}{2}\overline \psi \gamma^+\Sigma^3 \psi  +\overline{\psi}\gamma^+(\vec{\xi}\times i\vec{\partial})^3\psi
+ (\vec{E}\times \vec{A})^3+  E^i(\vec{\xi}\times \vec{\partial})^3A^i\right] \ ,
\label{freeop}
\end{eqnarray}
where $A^+=(A^0 + A^3)/\sqrt{2}$, and $E^i = F^{+i}$, and each term has a clear
physical interpretation. When all fields expressed in terms of creation and annihilation operators in light-cone
quantization, the above expressions involve just the ``good" components of the parton
fields. In Ref. \cite{Ji:1995cu}, the angular momentum evolution equation
was derived. The related angular momentum sum rule has been often quoted in the literature,
\begin{equation}
\frac{1}{2}=\frac{1}{2}\Sigma_q+ l_q+\Delta G+ l_g \ ,
\label{16}
\end{equation}
where all quantities are defined as the matrix elements of the above operators
in the nucleon helicity state. However, only the first term is gauge
invariant, whereas all other three terms are not. Since we are interested in
the parton physics, it is the most natural way to fix the light-cone gauge $A^+=0$,
which is what we will do in this paper.

The above canonical angular momentum operator has motivated introducing light-cone
AM density by H\"agler and Sch\"afer \cite{Hagler:1998kg} as generalized
angular momentum operators, and similarly by A. Harindranath and R. Kundu \cite{Harindranath:1998ve},
by Bashinsky and Jaffe \cite{Bashinsky:1998if}
\begin{equation}
l_q (x) =
\frac{1}{2\pi P^+}
\int d\lambda e^{i\lambda x}
\langle PS|\overline{\psi}(0)
\gamma^+ i(\xi^1\partial^2-\xi^2\partial^1)
\psi(\xi)|PS\rangle
\end{equation}
where $\lambda = \xi^-P^+$ and the integration over $\xi_\perp$ is implicit. To
take into account the transverse coordinates, it can be calculated as the forward
limit of an off-forward matrix element,
\begin{eqnarray}
l_q(x)=\epsilon^{\alpha\beta}\frac{i\partial}{\partial q_{\perp\alpha}}\left|_{q_\perp=0}\left[\int \frac{d\xi^-}{2\pi}e^{ixp^+\xi^-}
\left\langle P'S\left| \overline{\psi}(0)\gamma^+ i \partial_\perp^\beta \psi(\xi^-)\right|PS\right\rangle\right] \right.\ ,
\end{eqnarray}
where $q_\perp=P'-P$, $\alpha$ and $\beta$ only cover the transverse dimensions. Clearly, $\int dx l_q(x) = l_q$.

Likewise, the gluon helicity and OAM distributions can be defined in light-cone gauge as
\begin{eqnarray}
 \Delta g(x)  & =& \frac{1}{4\pi}\epsilon_{\alpha\beta} \int d\lambda
 e^{i\lambda x} \langle PS|
 F^{+\alpha}(0)A^\beta(\lambda n)|PS\rangle \label{19} \\
 l_g(x)  &=& \frac{1}{4\pi} \int d\lambda
 e^{i\lambda x} \langle PS|
 F^{+\alpha}(0)(\xi^1\partial^2-\xi^2\partial^1)A_\alpha(\lambda n)|PS\rangle
\end{eqnarray}
These are the gluon helicity distribution and canonical gluon OAM distribution.
The gluon helicity sum rule is
\begin{equation}
\Delta G=\int dx \Delta g(x) \ ,
\end{equation}
and similarly, $\int dx l_g(x) = l_g$. Thus all the components of the
canonical angular momentum sum rule in Eq. (\ref{16}) now have partonic interpretation.

\subsection{Relations Between Two OAM Distributions}

From the covariant derivative $iD = i\partial - gA$, one can introduce
the gauge-dependent potential angular momentum
term $l_{\rm pot} = -g\int d^3\xi \overline{\psi}(\vec{\xi}\times \vec{A})^3\psi$. Thus,
$L_q = l_q + l_{q,\rm pot}$. Similarly, one can introduce the relevant
parton distribution through its moments ~\cite{Hatta:2011ku},
\begin{eqnarray}
l^n_{q,\rm pot}=\frac{-\epsilon^{\alpha\beta}}{(P^+)^n}\frac{i\partial}{\partial q_{\perp\alpha}}\left|_{q_\perp=0}\!\left[\left\langle P'S\left|\overline{\psi}(0)\gamma^+\!\! \frac{1}{n}\sum_{k=0}^{n-1}\!\! \left(iD^+\right)^{n-1-k}\!\!\!\!g A_\perp^\beta(0)\!\!\left(iD^+\right)^k\!\!\!\psi(0)\right|PS\right\rangle \right]\right.\
.
\end{eqnarray}
Again the above quantity is defined in the light-cone gauge. Obviously, the canonical
AM distribution plus the potential AM distribution yield the manifest
gauge-invariant AM density,
\begin{equation}
L_q(x) = l_q(x) + l_{q,\rm pot}(x) \ .
\end{equation}
The total quark angular momentum density contribution to the nucleon helicity is
\begin{eqnarray}
    \tilde J_q(x) &=& \frac{1}{2}\Delta\Sigma(x) + L_q(x) \nonumber \\
            &=& \frac{1}{2}\Delta\Sigma(x) + l_q(x) + l_{q, \rm pot}(x)
\end{eqnarray}
which differs from $J_q(x)$ defined from twist-two GPDs of Eq.(\ref{jqx}) by a twist-three
distribution.

There is no further decomposition of the gluon contribution
to the nucleon spin in a gauge invariant fashion. However,
by using the equation of motion, the gluon part AM density tensor
can be written as,
\begin{equation}
M_g^{+\alpha\beta}(\xi)=\left(F^{+\alpha }A^\beta-F^{+\beta}A^\alpha\right)\!\!
-F^{+i}\left(\xi^\alpha\partial^\beta-\xi^\beta\partial^\alpha\right)\!\!
A_i+g\overline{\psi}\gamma^+\left(\xi^\alpha A^\beta-\xi^\beta A^\alpha\right)\psi \ ,\label{amg}
\end{equation}
where we have dropped out a total derivative term.
From the above expression, we can see that
the total contribution is gauge invariant, but
not the individual terms~\cite{Hoodbhoy:1999dr}. Therefore, the individual
contributions are not measurable in principle.
However, one can define them in the light-cone gauge and demand the same result in all other gauges (called GIE),
and explore their measurability.
Thus, generalizing to the light-cone distributions in the light-cone
gauge, we find that the total gluon contribution to the nucleon helicity
can be written~\cite{Ji:1996ek},
\begin{equation}
\tilde J_g(x)=\Delta g(x)+l_g(x)-\sum_q l_{q, \rm  pot}(x) \ .
\end{equation}
which again differs from $J_g(x)$ by a twist-three GPDs.

Therefore, the total partonic angular momentum distribution
to the nucleon helicity can be written as
\begin{eqnarray}
    \tilde J_q(x) + \tilde J_g(x) &=& \frac{1}{2} \Delta \Sigma (x) + \sum_q L_q(x) + \tilde J_g(x) \nonumber \\
                      &=& \frac{1}{2} \Delta \Sigma (x) + \Delta g(x) + \sum_q l_q(x)
                       + l_g(x)
\end{eqnarray}
when integrated over $x$, one gets both the sum rule in Eq. (1) and
the sum rule in Eq. (16).

\section{Probing Orbital Angular Momentum Distributions}

In this section, we consider the experimental probes of the OAM distributions. To this
effort, we relate them to experimentally measurable distributions such as
GPDs and Wigner distributions. In the process, one shall see that
the OAM distributions defined in the previous section require
twist-three processes to measure. It is a general rule of thumb in high-energy
scattering that the higher-twist distributions are more difficult to
probe than the leading twist ones. We will not go into the details
of specific experiments other than quote the possible 
hard-scattering processes. We note that the total quark
OAM can be measured in the leading-twist processes because of
Lorentz symmetry, which states that the fraction of the angular
momentum carried by quarks is independent of the polarization.

For the canonical angular momentum distributions in the light-cone gauge,
one has to find the corresponding gauge-invariant quantities
that are measurable in experiments. Thus in this section, we first consider the
so-called GIE of the light-cone-gauge quantities to identify the proper
observables. We then consider definitions of the twist-three GPDs and
their relations to the OAM distributions. Finally, we consider the relations
with Wigner distributions, exploring the possibility of obtaining
the OAM distributions through the quantum phase-space distributions.

\subsection{Gauge-Invariant Extension}

In the following discussion, the canonical AM operators, including the gluon spin operator,
are defined in the light-cone gauge. However, the experimental observables must be gauge-invariant.
To reconcile the difference, we introduce the concept of gauge-invariant extension
in the sense that these gauge-dependent operators in any other gauge must yield the same matrix
elements in the light-cone gauge. Thus for the example, the gluon helicity
distribution $\Delta g(x)$ in Eq.~(\ref{19}) has the following gauge-invariant form ~\cite{Manohar:1990jx},
\begin{eqnarray}
\Delta g(x)=\frac{i}{xP^+}\int\frac{d\xi^-}{2\pi} e^{ixp^+\xi^-}
\left\langle PS\left| F^{+i}(0)L_{[0,\xi^-]}\tilde F^{i+}(\xi^-)\right|PS\right\rangle \  ,
\end{eqnarray}
where $\tilde F^{+i}=\epsilon^{ij}F^{+j}$, and $L$ is the light-cone gauge link. Clearly, its first moment is no longer
a local operator. However, it reduces to the gluon spin operator in the light-cone
gauge. $\Delta g(x)$ will appear in the polarized structure functions
measured in deep inelastic scattering and the longitudinal
double spin asymmetries in $pp$ collisions. The experimental investigation
of this distribution is actively pursued at the relativistic heavy-ion collider
at the Brookhaven National Laboratory~\cite{deFlorian:2008mr}, and will be the main focus
in the planed electron ion collider in the near future~\cite{Boer:2011fh}.

If we introduce the following GIE of the partial derivative in light-cone gauge
\begin{equation}
   i\tilde{\partial}^\alpha_\perp = iD^\alpha_\perp(\xi) + \int^{\xi^-} d\eta^-  L_{[\xi^-,\eta^-]} gF^{+\alpha}(\eta^-,\xi_\perp) L_{[\eta^-,\xi^-]} \  ,
\end{equation}
where $L$ is the light-cone gauge link (the only ambiguity is the boundary condition
for the gluon potential at infinity which can be fixed by residual gauge definition, we
ignore this point in our discussion, see~\cite{Burkardt:2012sd})
we immediately find the canonical quark OAM distribution $l_q(x)$
is now gauge invariant, so is the canonical gluon OAM distribution $l_g(x)$.
The potential AM distribution involves $A^\alpha$, which can be made
gauge invariant through,
\begin{equation}
    \tilde{A}_\perp^\alpha(\xi) = \int^{\xi^-} d\eta^-  L_{[\xi^-,\eta^-]} gF^{+\alpha}(\eta^-,\xi_\perp) L_{[\eta^-,\xi^-]} \  . \label{aperp}
\end{equation}
Therefore all quantities defined in the light-cone gauge are now gauge
invariant although they are now highly non-local because of the light-cone
gauge links.

Through GIE,  we find the following relation between $l_q(x)$, $L_q(x)$ and $l_{q,\rm pot}$,
\begin{equation}
  l_q(x)= L_q(x) - l_{q,\rm pot}(x) \ .
\end{equation}
Similarly, we can define the gluon gauge-invariant OAM distribution through its moments
\begin{equation}
 L_g ^n = \frac{\epsilon^{\alpha\beta}}{4\pi (P^+)^n} \frac{i\partial}{\partial q_{\perp\alpha}} \left|_{q_{\perp}=0}\!\left[\left\langle\! P'S\left|\frac{1}{n}\!\sum_{k=0}^{n-1} F^{+i}(0)\!\left(iD^+\right)^{n-1-k}iD^\beta_\perp
 \left(iD^+\right)^k\! A^i(0)\right|PS\!\right\rangle\right] \right. \ ,
\end{equation}
and gluon potential AM distribution,
\begin{equation}
 l_{g,\rm pot}^n  = \frac{-\epsilon^{\alpha\beta}}{4\pi (P^+)^n}\frac{i\partial}{\partial q_{\perp\alpha}}\left|_{q_\perp=0}\!\left[
 \left\langle\! P'S\!\left|\frac{1}{n}\!\sum_{k=0}^{n-1}\!
 F^{+i}(0)\!\left(iD^+\right)^{n-1-k}\!\!\!gA^\beta(0)\!\left(iD^+\right)^k\! A^i(0)\!\right|\!PS\!\right\rangle\right]\right.\!\!\
 .
\end{equation}
It is also easy to see that
\begin{equation}
  l_g(x)= L_g(x) - l_{g,\rm pot}(x)\
  .
\end{equation}

\subsection{GPDs and OAM Distributions}

GPDs have been extensively discussed in the literature~\cite{Ji:1996nm,Diehl:2003ny,gpd3}.
In this section, we will focus on the twist-three GPDs
which are directly related to the various spin components derived
in the last section. We will also derive the connections between
various twist-three GPDs to illustrate the relations between different
terms in the spin sum rule.

In particular, we are interested in the twist-three GPDs associated
with the longitudinal polarized nucleon. Define a distribution with two light-cone fractions,
\begin{eqnarray}
&&\int\frac{d\lambda}{2\pi}\int\frac{d\mu}{2\pi}e^{i\lambda (x-y)}e^{i\mu y}\langle P' S|\overline{\psi}(0)
\gamma^+iD^\perp(\mu n)\psi(\lambda n)|PS\rangle\nonumber\\
&=&\frac{i\epsilon^{\perp\alpha}}{2}\Delta_\alpha H_D^{q(3)}(x,y,\eta,t)\overline{U}(P')\gamma^+\gamma_5U(P )+\cdots\ .
\label{hdq}
\end{eqnarray}
where $n$ is a conjugate vector $n^+=n_\perp=0$ with $n\cdot P=1$,
 $\eta$ is the skewness parameter, $t=\Delta^2$ with $\Delta=q=P'-P$.
It is straightforward to show that the moments of quark orbital angular
momentum distribution $L_q(x)$ is related to the moments of twist-three
GPDs in the forward limit\footnote{One may also define the OAM distribution
by $\int dy H_D^{q(3)}(x,y,0,0)$, which would correspond to the moment
definition in Eq.~(\ref{moment}) with $D^\perp$ only associated with
$\psi$ or $\bar \psi$ fields. The following discussions apply to this case
as well (see also the discussions in Ref.~\cite{hatta}).}
\begin{equation}
L_q^n=\int dx\int dy \frac{1}{n}\sum_{k=0}^{n-1}x^{n-1-k}(x-y)^k H_D^{q(3)}(x,y,0,0) \
.
\end{equation}

%
Similarly, we can define the following twist-three GPDs associated with
gluon which is related to $L_g(x)$,
\begin{eqnarray}
&&\int\frac{d\lambda}{2\pi P^+}\int\frac{d\mu}{2\pi}e^{i\lambda (x-y)}e^{i\mu y}\langle P' S|F^{+i}(0)
iD^\perp(\mu n) F^{+i}(\lambda n)|PS\rangle\nonumber\\
&=&\frac{i\epsilon^{\perp\alpha}}{4}\Delta_\alpha H_D^{g(3)}(x,y,\eta,t)\overline{U}(P')\gamma^+\gamma_5U(P )+\cdots\ .
\label{hdg}
\end{eqnarray}
Again, the moments of $L_g(x)$ is related to the moments of $H_D^{g(3)}(x,y,\eta,t)$ in the forward limit,
\begin{equation}
L^n_g=\int dx\int dy \frac{1}{n}\sum_{k=0}^{n-1} x^{n-1-k}(x-y)^{k-1} H_D^{g(3)}(x,y,0,0) \ .
\end{equation}
For $l_q$ and $l_g$, we have
\begin{eqnarray}
&&\int\frac{d\lambda}{2\pi}e^{i\lambda x} \left\langle P' S|\overline{\psi}(0)
\gamma^+i\tilde\partial^\perp\psi(\lambda n)|PS\right\rangle\nonumber\\
&=&\frac{i\epsilon^{\perp\alpha}}{2}\Delta_\alpha \tilde{H}_q^{(3)}(x,\eta,t)\overline{U}(P')\gamma^+\gamma_5U(P )+\cdots\ ,\\
&&\int\frac{d\lambda}{2\pi P^+}e^{i\lambda x} \langle P' S|F^{+i}(0)
i\tilde\partial^\perp F^{+i}(\lambda n)|PS\rangle\nonumber\\
&=&\frac{i\epsilon^{\perp\alpha}}{4}\Delta_\alpha \tilde H_g^{(3)}(x,\eta,t)\overline{U}(P')\gamma^+\gamma_5U(P )+\cdots\ .
\end{eqnarray}
We have,
\begin{eqnarray}
l_q(x)&=& \tilde H_q^{(3)}(x,0,0)\ ,\\
xl_g(x)&=&\tilde H_g^{(3)}(x,0,0)\ .
\end{eqnarray}
The associated potential OAM terms depend on the F-type twist-three GPDs. To discuss
the connections, we will start with more general forms for the
twist-three GPDs.

In addition to the $D$-type twist-three GPDs of Eqs.~(\ref{hdq},\ref{hdg}), there are also $F$-type
GPDs,
\begin{eqnarray}
&&{\int\frac{d\lambda}{2\pi}\int\frac{d\mu}{2\pi P^+}e^{i(x-y)\lambda}e^{iy\mu} \langle P' S|\overline{\psi}(0)
\gamma^+\!gF^{+\perp}(\mu n)\psi(\lambda n)|PS\rangle}\nonumber\\
&=&\frac{\epsilon^{\perp\alpha}}{2}\Delta_\alpha H_F^{q(3)}(x,y,\eta,t)\overline{U}(P')\gamma^+\gamma_5U(P )+\cdots\ ,\\
&&{\int\frac{d\lambda}{2\pi P^+}\int\frac{d\mu}{2\pi P^+}e^{i(x-y)\lambda}e^{iy\mu}
 \langle P' S|F^{+i}(0)gF^{+\perp}(\mu n)F^{+i}(\lambda n)|PS\rangle}\nonumber\\
&=&\frac{\epsilon^{\perp\alpha}}{4}\Delta_\alpha H_F^{g(3)}(x,y,\eta,t)\overline{U}(P')\gamma^+\gamma_5U(P )+\cdots\ .
\end{eqnarray}
It is the F-type twist-three GPDs that are related to the potential OAMs for the quarks and gluons,
respectively,
\begin{eqnarray}
l^n_{q,\rm pot}&=&-\int dx \int dy \frac{1}{n}\sum_{k=0}^{n-1} x^{n-1-k}(x-y)^k P\frac{1}{y} H_F^{q(3)}(x,y,0,0)\ , \\
l^n_{g,\rm pot}&=&-\int dx \int dy \frac{1}{n}\sum_{k=0}^{n-1} x^{n-1-k}(x-y)^{k-1} P\frac{1}{y} H_F^{g(3)}(x,y,0,0)\ .
\end{eqnarray}

Similar to that for the twist-three quark-gluon-quark correlation function in the forward
limit, the D-type and F-type twist-three GPDs are related to each other~\cite{Eguchi:2006qz,Zhou:2009jm},
\begin{equation}
{H_D^{q,g(3)}(x,y,0,0)=-P\frac{1}{y}H_F^{q,g(3)}(x,y)+\delta(y)\tilde H_{q,g}^{(3)}(x,0,0)} \ ,
\end{equation}
respectively. These relations are consistent with the relations,
\begin{equation}
L_{q/g}(x)=l_{q/g}(x)+l_{q/g, \rm pot}(x)
\end{equation}
in the previous sections.

\subsection{Wigner Distribution and Parton OAM Distributions}

Parton Wigner distribution was introduced in Ref.~\cite{Belitsky:2003nz} to unify the
transverse momentum dependent distributions (TMDs) and the GPDs. They
describe the phase space distribution of partons in nucleon.
In particular, they contain information on momentum and coordinate
dependence in the transverse plane perpendicular to the
nucleon momentum direction. Therefore, the Wigner distribution
will naturally provide the spin-orbital correlation
which is important to extract the parton orbital angular momentum.
This has been demonstrated explicitly in recent studies~\cite{Ji:2012sj,Lorce:2011kd,Hatta:2011ku}.

We define the Wigner distribution for the quark~\cite{Belitsky:2003nz},
\begin{equation}
W^q_\Gamma(x,\vec k_\perp,\vec{r})=\int \frac{d\eta^-d^2\vec \eta_\perp}{(2\pi)^3}e^{ik\cdot \eta}
\langle P|
\overline{\Psi}(\vec{r}-\frac{\eta}{2})\Gamma \Psi(\vec{r}+\frac{\eta}{2})|P\rangle \ ,
\end{equation}
where $x$ represents the longitudinal momentum fraction carried by the quark,
$k_\perp$ the transverse momentum, $\vec{r}$ the coordinate space variable, and $\Gamma$
the Dirac matrix to project out the particular quark distribution.
The quark field $\Psi$ contains the relevant gauge link to guarantee the gauge invariance
of the above definition.

If we further integrate over $r_z$, we will obtain the transverse Wigner distribution,
which can be interpreted as the phase space ($x$,$\vec k_\perp$,$\vec b_\perp \equiv=r_\perp$) distribution of the parton
in the transverse plane perpendicular to the nucleon momentum direction.
We are interested in obtaining the Wigner distribution
for the quark in a longitudinal polarized nucleon.
With the correct Dirac matrix projection, we define the following
phase space distribution,
\begin{equation}
W^q(x,\vec k_\perp,\vec b_\perp)=\int\!\!\!\frac{d^2\vec q_\perp}{(2\pi)^2}e^{-i\vec q_\perp \cdot \vec b_\perp}\!\!\!\int \frac{d\eta^-d^2\vec\eta_\perp}{(2\pi)^3}e^{ik\cdot \eta}
\langle P+\frac{\vec q_\perp}{2}|
\overline{\Psi}(-\frac{\eta}{2})\gamma^+ \Psi(\frac{\eta}{2})|P-\frac{\vec q_\perp}{2}\rangle \ ,
\end{equation}
where $\vec b_\perp$ and $\vec k_\perp$ are transverse coordinate and
momentum variables, respectively.

Because of the transverse momentum dependence, the above introduced
Wigner distributions involve the gauge links which are process dependent.
For the relevance of the quark OAM from the quark Wigner distributions,
it was found two options can be chosen,
\begin{eqnarray}
        \Psi_{LC}(\xi)&=&{{P}}\left[\exp\left(-ig\int^\infty_0 d\lambda n \cdot A(\lambda n+\xi)\right)\right]   \psi(\xi) \\
       \Psi_{FS}(\xi)&=&{{P}}\left[\exp\left(-ig\int^\infty_0 d\lambda \xi \cdot A(\lambda\xi)\right)\right] \psi(\xi)  \ .
\end{eqnarray}
The light-cone gauge link in $ \Psi_{LC}(\xi)$ is appropriate for high energy process,
such as semi-inclusive hadron production in deep inelastic
scattering~\cite{Ji:2004wu} (see also~\cite{Burkardt:2012sd} about 
the boundary conditions at infinity). The second choice $ \Psi_{FS}(\xi)$ is a straightline
gauge link along the direction of spacetime position $\xi^\mu$.
This link reduces to unity in the Fock-Schwinger gauge, $\xi\cdot A(\xi)=0$.

It is straightforward to show that the straightline gauge link corresponds
to the gauge invariant OAM for the quark,
\begin{eqnarray}
L_q(x)   =  \int { (\vec{b}_\perp \times \vec{k}_\perp)^3 W_{FS}^q (x, \vec{b}_\perp,  \vec{k}_\perp) d^2\vec{b}_\perp d^2 \vec{k}_\perp} \ ,
\label{sum}
\end{eqnarray}
which gives a parton picture for the gauge-invariant OAM $L_q(x)$.
For the canonical quark OAM $ l_q(x)$, we have,
\begin{equation}
  l_q(x) = \int  (\vec{b}_\perp \times \vec{k}_\perp)^3 W^q_{LC}(x, \vec{b}_\perp,  \vec{k}_\perp)  {{d^2\vec{b}_\perp d^2 \vec{k}_\perp}} \ .
\end{equation}
From the above equations, we find that the gauge invariant OAM $L_q(x)$
and the canonical OAM $l_q(x)$ are unified as projection of the
associated quark Wigner distribution with two different choices for
the gauge link.

\subsection{Wigner Distributions for the Gluons}

Similarly, we can define the Wigner distribution for the gluons,
\begin{equation}
xW^g(x,k_\perp,{b_\perp})=\int\!\!\frac{d^2q_\perp}{(2\pi)^2}e^{-iq_\perp\cdot b_\perp}\!\!\int\!\!\frac{d\eta^-d^2\eta_\perp}{(2\pi)^3}e^{ik\cdot \eta}
\langle P+\frac{q_\perp}{2}\left|{\bf F}^{+i}(-\frac{\eta}{2}) {\bf F}^{+i}(\frac{\eta}{2})\right|P-\frac{q_\perp}{2}\rangle \ ,
\end{equation}
where ${\bf F}$ contains the relevant gauge links,
\begin{eqnarray}
{\bf F}_{FS}^{+i}(\xi)&=&{{{P}}\left[\exp\left(-ig\int^\infty_0 d\lambda \xi \cdot A(\lambda\xi)\right)\right] F^{+i}(\xi)}\\
{\bf F}_{LC}^{+i}(\xi)&=&{{{P}}\left[\exp\left(-ig\int^\infty_0 d\lambda n \cdot A(\lambda n+\xi)\right)\right]   F^{+i}(\xi)} \ .
\end{eqnarray}
Here, the gauge links are in the adjoint representations. We have chosen the future
pointing gauge links. We notice that there
are different choices for the gauge links associated with the transverse momentum
dependent gluon distributions (see, e.g., Ref.~\cite{Dominguez:2010xd}).
Future pointing gauge link corresponds to the processes involving
only final state interactions associated with the gluon distribution,
such as dijet-correlation in DIS~\cite{Dominguez:2010xd}. Because
of the time-reversal invariance, for the above quantity, the future pointing
gauge link yields the same Wigner distribution as that for backward pointing
gauge link in the process such as Higgs boson production in $pp$ collisions.
Following the above calculations, we find that the gluon OAM can be
constructed from these Wigner distributions~\cite{Lorce:2011kd},
\begin{eqnarray}
 L_g(x)&=&\int  (\vec{b}_\perp \times \vec{k}_\perp)^3 W_{FS}^g(x, \vec{b}_\perp,  \vec{k}_\perp)  {{d^2\vec{b}_\perp d^2 \vec{k}_\perp}} \\
l_g(x)&=&\int  (\vec{b}_\perp \times \vec{k}_\perp)^3 W_{LC}^g(x, \vec{b}_\perp,  \vec{k}_\perp)  {{d^2\vec{b}_\perp d^2 \vec{k}_\perp}} \ .
\end{eqnarray}
In high energy processes, these Wigner distributions could be measurable
in hard exclusive processes.

\section{Summary}

In conclusion, we have studied the parton sum rule for the spin of 
a longitudinal polarized nucleon, in terms of the parton helicity
contributions and OAM contributions. We have identified the
twist-three GPDs as direct probes for the parton OAMs.
We also demonstrated that the quark and gluon OAMs can be
related to the quantum phase space distributions.
In particular, the canonical and gauge-invariant
quark OAMs correspond to the different choices for the
associated gauge links structure in the Wigner distributions.

Since the twist-three GPDs can be used to probe the parton
OAMs directly, we can now in principle measure the different components
of the spin sum rule. These GPDs shall be studied in the
hard exclusive processes.
However, the associated spin asymmetries always contain
the leading-twist contributions as in, for example, deeply virtual
Compton scattering process, it is not easy to extract these
twist-three GPDs. We have to make more detailed studies
on these processes and learn how to extract the relevant
distributions.

Moreover, the Wigner distributions shall also be accessible
through high energy processes. We will investigate these issues
in the future publications.

We thank Y.~Hatta for the discussions and communications on their recent work~\cite{hatta}.
This work was partially supported by the U.
S. Department of Energy via grants DE-FG02-93ER-40762
and DE-AC02-05CH11231 and a grant from National Science Foundation of China (X.J.).

\end{document}